\documentclass[twocolumn]{revtex4}
\usepackage{amsfonts,amsmath,amssymb,latexsym,epsfig,hyperref}
\hypersetup{
    colorlinks,%
    citecolor=blue,%
    filecolor=blue,%
    linkcolor=blue,%
    urlcolor=blue
}

\newcommand{\pr}[1]{\left( #1\right)}
\newcommand{\prr}[1]{\left[ #1 \right]}
\newcommand{\es}[1]{\begin{equation}\begin{split}#1\end{split}\end{equation}}
\newcommand{\est}[1]{\begin{equation*}\begin{split}#1\end{split}\end{equation*}}
\newcommand{\R}{\mathbb{R}}

\newcommand{\V}{\mathcal{V}}
\newcommand{\rr}{\mathbf{r}}

\begin{document}
\title{k-connectivity for confined random networks}
\author{Orestis Georgiou$^{1,2}$, Carl P. Dettmann$^{2}$, Justin P. Coon$^{1,3}$}

\affiliation{1 Toshiba Telecommunications Research Laboratory, 32 Queens Square, Bristol, BS1 4ND, UK.}
\affiliation{2 School of Mathematics, University of Bristol, University Walk, Bristol BS8 1TW, UK.}
\affiliation{3 Department of Electrical and Electronic Engineering, University of Bristol, BS8 1UB, Bristol, UK.}

\begin{abstract}
$k$-connectivity is an important measure of network robustness and resilience to random faults and disruptions.
We undertake both local and global approaches to $k$-connectivity and calculate closed form analytic formulas for the probability that a confined random network remains fully connected after the removal of $k-1$ nodes.
Our analysis reveals that $k$-connectivity is governed by microscopic details of the network domain such as sharp corners rather than the macroscopic total volume.
Hence, our results can aid in the design of reliable networks, an important problem in e.g. wireless ad hoc and sensor networks.
\end{abstract}

\maketitle

\section{Introduction}
\label{sec:intro}

\begin{figure}[t]
\begin{center}
\includegraphics[scale=0.2]{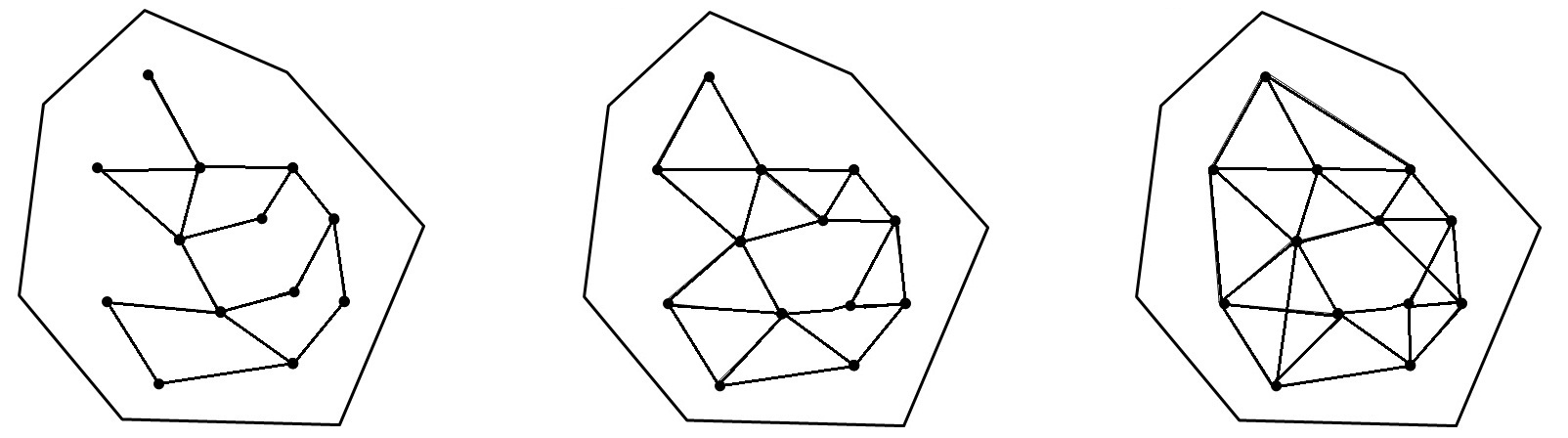}
\caption{\label{fig:123conn} Examples of networks with $N=13$ nodes in a convex domain satisfying $k=1,2$, and $3$ connectivity from left to right.}
\end{center}
\end{figure}

Random geometric networks~\cite{penrose2003random} consist of a collection of nodes randomly scattered in a region of space, pairwise connected with a relative position dependent probability.
An important application of network theory (amongst many others) is in wireless communications. Communication networks, provide rapid transfer of information across space, with applications ranging from the Internet, tele-medicine, intelligent transport, tracking of endangered species, hazard detection systems, security monitoring, etcetera (see Refs.~\cite{haenggi2009stochastic,li2009connectivity,wang2009understanding} and references therein).
Consequently, resilience to random faults or attacks are of paramount importance for the smooth functionality of the system. 
A typical measure of network robustness (particularly for communication networks) is $k$-connectivity, that is, if any $k-1$ nodes are randomly chosen and removed the remaining network remains fully connected.
Equivalently, a network is said to be $k$-connected if for each
pair of nodes there exist at least $k$ mutually independent paths connecting them~\cite{penrose2003random}.
Fig.~\ref{fig:123conn} shows examples of $k=1,2$, and $3$ connected networks.
In general, the random removal of nodes may result from a technical failure (e.g. a software/hardware malfunction) or a random attack that may disrupt network functionality and lead to cascades of catastrophic failures \cite{buldyrev2010catastrophic}.

Historically, the classical problem of $k$-connectivity has been addressed in the asymptotic limit of infinite network size and deterministic link formation whenever nodes are within a certain range \cite{almasaeid2009minimum,wan2010asymptotic}.
Instead, we consider spatially \textit{confined} networks formed by \textit{probabilistic} link connections.
Many real networks, although large, are not of infinite size and are often confined within a finite region. 
This makes classic percolation \cite{Bela} and random graph \cite{penrose2003random} approaches undesirable or even unsuitable in certain occasions.
Furthermore, noise, uncertainty, or the variation in connectivity range of individual nodes justifies the use of probabilistic link formations rather than deterministic ones.
In fact, probabilistic link-models are much preferred in many applications, for instance in wireless communications where they can adequately account for small-scale scattering and fading effects \cite{tse2005fundamentals}.

In this letter, we derive closed form analytic formulas for the probability of a random network residing in arbitrary 2 and 3 dimensional convex domains to be $k$-connected. 
We achieve this by undertaking both \textit{local} and \textit{global} approaches to $k$-connectivity.
The former refers to the local perspective of a single node, while the latter to the global perspective of large clusters of nodes.
Contrary to the expected universal features of large networks, our analysis reveals that the global network observable of $k$-connectivity is governed by distinct microscopic details of the network boundary such as sharp corners rather than macroscopic ones such as the domain volume.
Significantly, our analytic results provide sufficient quantitative detail to support optimization of system parameters in order to design reliable networks \cite{yun10}, mitigate boundary effects and avoid the need for heavy computer simulations.
Finally, the techniques presented below provide a flexible and mathematically tractable framework for further analysis of confined random networks.

%%%%%%%%%%%%%%%%%%%%%%%%%%

\section{Description of the problem}
\label{sec:def}

The networks we wish to model consists of $N$ randomly distributed nodes with locations ${\bf r}_i\in{\cal V}$ a convex subset of $\mathbb{R}^d$, with $i=1,2,\ldots N$, according to a uniform
density $\rho=N/V$, where $V=|{\cal V}|$ and
$|\cdot|$ denotes the size of the set using the Lebesgue measure of the
appropriate dimension or the cardinality of a finite set.
We consider convex geometries as we only allow for line-of-sight links between nodes.
After deployment of the nodes in $\cal V$, communication links between pairs of nodes are established with probability $H(r_{ij})$, often written as $H_{ij}$ where
$r_{ij}=|{\bf r}_j-{\bf r}_i|$ is the distance between nodes $i$ and $j$.
Hence, the relevant network $g=(S,L)$ is formed, consisting of the set of nodes $S=\{1,2,3,\ldots,N\}$ paired
by the collection $L\subseteq \{(i,j)\in S^2 :i<j\}$
of direct links.

We maintain physical relevance by adopting a specific pair connectedness function $H_{ij}$ derived from wireless communication theory~\cite{tse2005fundamentals} and applicable to ad hoc and sensor networks.
In particular, we use a Rayleigh fading model suitable when there is no dominant communication channel along a single line of sight between transmitter and receiver but rather a sum of many paths with randomised phases.
The resulting connection probability between two nodes a distance $r$ apart is given by
\es{
H(r)= e^{-\beta r^{\eta}}
\label{H},}
where $\beta$ depends on for example the transmission wavelength, signal power, etc., and sets the characteristic connection length $r_0=\beta^{-1/\eta}$ and the parameter $\eta$ is called the path loss exponent and is typically set to $\eta=2$ corresponding to propagation in free space but is experimentally observed to be $\eta>2$ for cluttered environments e.g. heavily built-up urban environments \cite{tse2005fundamentals,Seidel92}.
Unless otherwise stated, we will use $\eta=2$. In doing so $H(r)$ is a Gaussian function thus rendering the mathematics tractable.
It is worth noting that in the limit of $\eta\rightarrow \infty$, the connection between nodes is no longer probabilistic and converges to the well studied case in geometric graph theory~\cite{penrose1999k}, the \textit{unit disk} model with an \textit{on/off} connection range at $r_0$.
However, much less effort has been dedicated on the connectivity properties of networks formed by probabilistic (or soft) connectivity functions $H(r)$.

The connectivity of the network can be measured by checking whether any node $i$ can communicate in a multi-hop fashion with any other node $j\not= i$. If this is the case then the network is said to be fully connected, or $1$-connected.
In computer simulations, one initiates a search algorithm to count the number of connected components (clusters). If only a single cluster is found then $1$-connectivity is established.
Typically such algorithms have computation complexity of $\mathcal{O} (N \ln N)$.
For $2$-connectivity, a random node is removed from the network and the search algorithm is run. If successful, the node is replaced and its original links reconnected and a different node is removed and the algorithm is repeated. If successful for all $N$ nodes then $2-$connectivity is established. Therefore, the computational complexity is now of $\mathcal{O} \pr{ N(N-1)\ln (N-1)}$.
For $k$-connectivity this number grows like $\sim N^{k}(\ln N)$.

For a given node density $\rho$ we are interested in the probability $P_{fc}(k)$ of a randomly formed network to be $k$-connected. Of course this depends on the connectivity function $H_{ij}$ and the domain shape. In order to produce the $S$-curve describing $P_{fc}(k)$ as a function of the density $\rho$, one needs to perform a Monte Carlo computer simulation, averaging over many realizations of the above described algorithm. In this paper we will provide closed form analytical formulas which accurately predict this function in $2$ and $3$ dimensional convex domains hence eliminating the need for such heavy computer simulations.

%%%%%%%%%%%%%%%%%%%%%%%%%%%%%%%%%%%%%%

\section{Local Approach}
\label{sec:local}

We adopt a \textit{bottom-up} approach and begin our investigation from the local view point of a single node.
The probability that node $i$ situated at $\rr_i$ connects with a randomly chosen node $j$ is obtained by averaging over all possible node positions $\rr_j\in\V$
\es{
H_{i}(\textbf{r}_i)=\frac{1}{V}\int_{\mathcal{V}} H(r_{ij}) \textrm{d} \textbf{r}_{j}
.
\label{Hi}}
Note that $H_i$ can be studied in detail for different pair-connectedness functions $H(r)$ (e.g. anisotropic), and power and diversity scaling laws can be deduced through in-depth analysis~\cite{coon2012full,coonconnectivity}.
In the discussion that follows however, we will mainly be concerned with global network observables and concentrate on how local properties of nodes contribute to them.

Since nodes are deployed independently with a uniform density, the probability that node $i$ connects with exactly $k$ other nodes (i.e. node $i$ is of degree $k$) denoted here by $d_i(k)$, is given by the binomial distribution
\es{
d_i(k)= \binom{N-1}{k} H_i^{k}(1-H_i)^{N-1-k}
,
\label{bin}}
which for $H(r)$ a hard step function ($\eta=\infty$) would correspond to the probability of finding $k$ nodes (excluding node $i$) in the $r_0$-neighbourhood of $\mathbf{r}_i$, and $(N\!-\!1\!-\!k)$ nodes elsewhere in the available domain $\mathcal{V}$.
Another way of expressing~\eqref{bin} is achieved by noting that if $N$ is large and $H_{i}$ is small, $d_i(k)$ is well approximated by the Poisson distribution
\es{
d_i(k)\approx \frac{\lambda_{i}^{k}}{k!} e^{-\lambda_i}, \qquad D_{i}(k)=\sum_{m=0}^{k}d_{i}(m)
\label{nd},}
where $\lambda_i= (N-1)H_i$, and $D_{i}(k)$
is the corresponding cumulative distribution function.
This approximation is justified here as $V\gg 1$, thus making $H_{i}\ll1$ and $N\gg1$.

%%%%%%%%%%%%%%%%%%%%%%%%%%%%%%%

\begin{figure}[t]
\begin{center}
\includegraphics[scale=0.22]{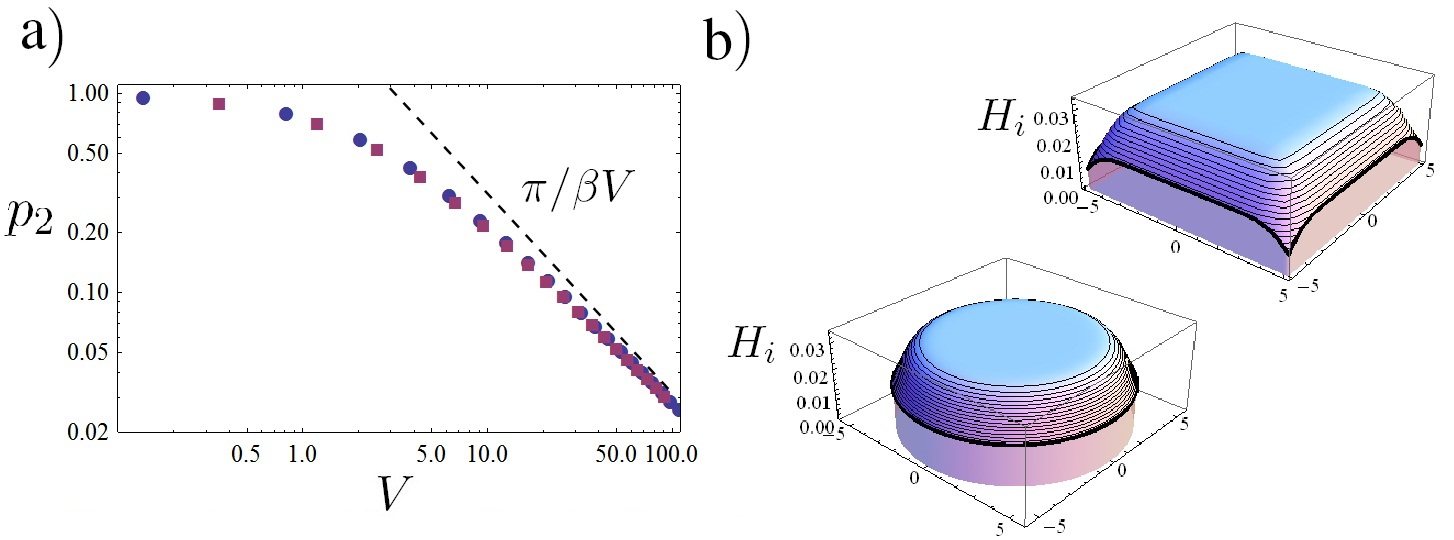}
\caption{\label{fig:pair} a) Log-log plot of $p_2$ vs $V$ for square and circular disk domains of equal volumes illustrated with square/disk markers using $\beta=1$. The dashed line is the asymptotic distribution $\pi/(\beta V)$.
b) 3D plots of $H_i(\textbf{r}_i)$ for square and circular disk domains both of equal volumes $V=100$.}
\end{center}
\end{figure}

Integrating  \eqref{Hi} over $\textbf{r}_{i}$
gives $p_{2}$, the probability that two randomly selected nodes connect to form a pair
\es{
p_{2} = \frac{1}{V^2}\int_{\mathcal{V}^{2}} H(r_{ij}) \textrm{d} \textbf{r}_{i} \textrm{d} \textbf{r}_{j}
\label{pair}.}
Numerical integration of \eqref{pair} for a square and a circular disk domain of equal volumes reveals that $p_{2}$ is very much insensitive to the domain's shape and decays like $\sim\pi/\beta V$ as $V\rightarrow\infty$ (see Fig.~\ref{fig:pair} a)).
Note that the important length scale here is not the aspect ratio of the domain but the ratio of the connection range $r_0$ and the typical size of the system.  
For example, $p_2$ of an elongated rectangular domain would decay as predicted above if its shortest side is much greater than $r_0$ and its total volume $V$ is large.

The average number of nodes connected to a node in a network is called the \textit{mean degree}. From the degree distribution~\eqref{nd} we can immediately deduce that the mean degree (as well as the variance) is just $\lambda=\int \lambda_i  \textrm{d}\textbf{r}_i /V= (N-1) p_2$ and is also highly insensitive to the domain shape.
Moreover, for $N$ and $V$ large, we have that $\lambda \sim (N-1)\pi/(\beta V)\approx \rho \pi/\beta$.

Finally, we turn to investigate short range correlations and look at the 2-point correlation function in an infinite domain. Here, we keep $\eta$ general and consider 3 nodes with polar coordinates $(r_1,\theta_1)$, $(0,0)$ and $(r,0)$ and define the two point correlation function as
\es{
C(r,\eta)=\frac{\int_{\R^2} H_{12}H_{13} \textrm{d}\textbf{r}_1}{\int_{\R^2} H_{12}\textrm{d}\textbf{r}_1}
.
\label{cor}}
Equation \eqref{cor} is nothing more than Bayes' theorem measuring how likely it is that node 1 connects with node 3, given that node 1 is connected with node 2.
The denominator of \eqref{cor} amounts to $2\pi \Gamma(2/\eta) /\eta \beta^{2/\eta}$. Expanding the integrand of the numerator in powers of $r\ll 1$ we obtain \es{C(r,\eta)=2^{-2/\eta} - \frac{\eta \beta^{2/\eta}}{8 \Gamma\pr{1+ 2/\eta}}r^{2} +\mathcal{O}(r^{4})
.
\label{cor2}}
Note that for $H(r)$ a deterministic step function (i.e. $\eta=\infty$) equation \eqref{cor2} diverges and instead is given by the circle-circle intersection $C(r,\infty)=1- 2r/(\pi r_0) + \mathcal{O}(r^{3})$.
This short calculation suggests that two nearby nodes are strongly correlated for deterministic (hard) connectivity functions with $C(0,\infty)=1$. This is not the case however for probabilistic (soft) connectivity functions. Indeed for the softest case of $\eta=2$ we have $C(0,2)=1/2$. It is reasonable to expect that this distinction of hard versus soft $H(r)$ persists in higher dimensions and for correlations between $n\!>\!2$ nodes. 

%%%%%%%%%%%%%%%%%%%%%%%%

\section{Minimum Network Degree}
\label{sec:mindeg}

For a given configuration of node positions we define $P_{md}(\mathbf{r}_{1},\ldots \mathbf{r}_{N},k)$ as the probability of the corresponding network to have a minimum degree of at least $k$, i.e. every node is connected to at least $k$ other nodes. The average of this quantity over all possible configurations $P_{md}(k)=\langle  P_{md}(\mathbf{r}_{1},\ldots \mathbf{r}_{N},k) \rangle$ is the overall probability of a network with minimum degree $k$.
We define the spatial average of an observable $O$ over all possible node configurations as
\es{\langle O\rangle = \frac{1}{V^{N}}\int_{\mathcal{V}^{N}} O(\mathbf{r}_{1},\mathbf{r}_{2},\ldots, \mathbf{r}_{N}) \mathrm{ d}\mathbf{r}_{1} \mathrm{ d}\mathbf{r}_{2}\ldots \mathrm{ d}\mathbf{r}_{N}
\label{average}
.}
Assuming that for $N\gg 1$, the degree of node $i$ is almost independent of the degree of node $j\not=i$, we write \cite{bettstetter2002minimum}
\es{
P_{md}(k)& = \langle\prod_{i=1}^{N} P(\textrm{degree}(\textbf{r}_{i})\geq k) \rangle =\langle \prod_{i=1}^{N} \pr{ 1- D_{i}(k-1)} \rangle \\
&\approx \prr{1- \langle D_{i}(k-1) \rangle}^{N}
\label{pmd}
.}
The seemingly strong assumption of independence made here is justified when performing the spatial average~\eqref{average} as two nodes $i$ and $j$ are sufficiently apart and thus uncorrelated for most node positions in $\mathcal{V}$. 
This is particularly true in dense networks and therefore we expect the approximation in \eqref{pmd} to improve as $N$ grows.
Furthermore, for soft connectivity functions $H(r)$, we expect only weak correlations between $n$-tuples of nodes, as argued in the previous section (see Eq.~\eqref{cor2}).

Substituting in~\eqref{pmd} the definition of $D_{i}(k)$, performing the average and omitting terms of $\mathcal{O}(1/N)$ we arrive at
\es{
P_{md}(k)\!\approx \!\prr{1-\! \sum_{m=0}^{k-1}\!\frac{\rho^{m}}{m!} \frac{1}{V}\int_{\mathcal{V}}\!\! M_{H}^{m} (\textbf{r}_i) e^{-\rho M_{H} (\textbf{r}_i)}
\textrm{d}  \textbf{r}_i}^{N}
\label{Pmd}}
where $M_{H} (\textbf{r}_i)=V H_{i}$.
There are several important observations to be highlighted here. Firstly, in the high density limit we expect the integrals appearing in equation~\eqref{Pmd} to be dominated by contributions where $M_{H}(\textbf{r}_i)$ is small. This is due to the exponential part of the integrand dominating over the power. Secondly, we expect that $M_{H}(\textbf{r}_i)$ is small near the domain  boundaries i.e. near the corners and edges (see Fig.~\ref{fig:pair} b)) which physically correspond to the most \textit{hard to connect to} regions of $\mathcal{V}$. We conclude by noting that unlike $\lambda$, $P_{md}(k)$ is strongly influenced by the details of the domain boundary.

%%%%%%%%%%%%%%%%%%%%%%%%%%%%%%%%%%%%%%%%

\section{General analytic formulas} 
\label{sec:general}

\begin{figure}[t]
\begin{center}
\includegraphics[scale=0.2]{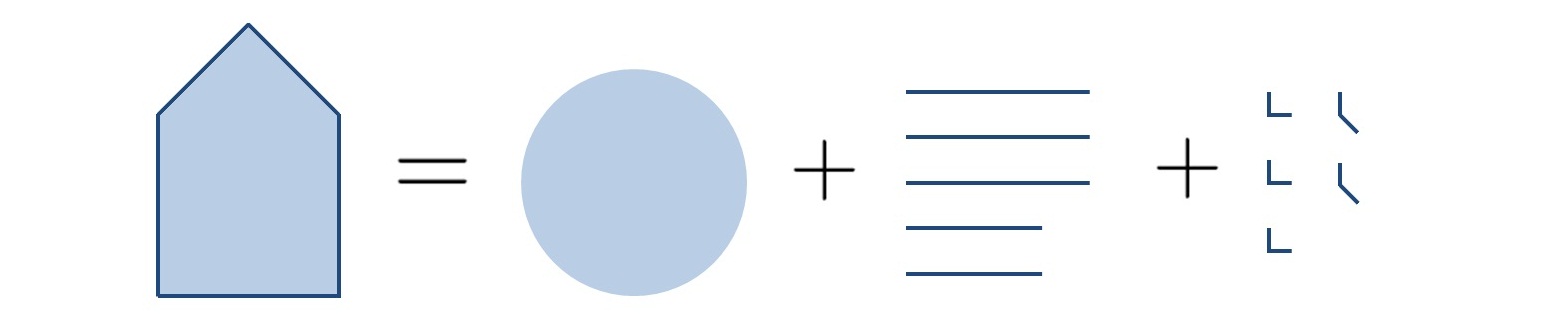}
\caption{\label{fig:house} Decomposition of a house domain into contributions from the bulk, the $5$ edges and $5$ corners.}
\end{center}
\end{figure}

The integrals to be approximated in \eqref{Pmd} are of the form
\es{
I_m = \int_{\mathcal{V}} M_{H}^{m} (\textbf{r}_2) e^{-\rho M_{H} (\textbf{r}_2)}
\textrm{d}  \textbf{r}_2 ,
\label{int}}
where $M_{H}(\textbf{r}_2)= \int_\mathcal{V} H_{12} \textrm{d} \textbf{r}_{1}$.
Due to the short-range interactions between nodes, hard to connect to regions become almost independent of each-other and so we approximate the integral of~\eqref{int} by a sum of independent contributions due to different boundary objects \cite{coon2012impact}.
That is, for an arbitrary convex domain $\mathcal{V}\subset \R^{2}$ the integral of \eqref{int} can be approximated by the sum of the bulk contribution ($B$), and a number of edge ($E$) and corner ($C$) contributions
\es{
I_m \approx I_m^{(B)} + \sum_{E_i} I_m^{(E_i)} + \sum_{C_i} I_m^{(C_i)}
\label{inta}
.}
For instance, the ``house" domain shown in Fig.~\ref{fig:house}, as far as $I_m$ is concerned, decomposes into a homogeneous bulk contribution, $5$ edges  and finally $5$ corners. 
The only restriction in this approximation is that the individual boundary elements are sufficiently apart (approximately a distance greater than $2 r_0$).

We now consider each of these contributions separately, first for $d=2$ and later for $d=3$. Due to length restrictions, we will omit any tedious calculations and give only important steps and final results.
The bulk contribution to~\eqref{inta} can be obtained by ignoring any boundary effects and thus considering a homogeneous domain leading to $M_{H}^{(B)} \! = \frac{\pi}{\beta}$,
such that %\eqref{int} gives 
$I_m^{(B)}=  V \pr{\frac{\pi}{\beta}}^{m}e^{-\rho \frac{\pi}{\beta}}$.

The edge contribution to~\eqref{inta} can be obtained by ignoring any curvature effects and thus considering the positive half plane $\mathcal{V}= [0,\infty) \times (-\infty,\infty)$ entailing
\est{
M_{H}^{(E)}(x_{2}) 
&= \int_{-\infty}^{\infty}\! \int_{0}^{\infty} e^{-\beta\pr{(x_1-x_2)^2+(y_1-y_2)^2}} \textrm{d} x_1 \textrm{d} y_1 \\
&= \frac{\pi}{2\beta}+ \sqrt{\frac{\pi}{\beta}} x_{2} + \mathcal{O}(x_2^3),}
\es{
I_m^{(E)}  
&=\int_{0}^{L}  \int_{0}^{\infty}  (M_{H}^{(E)}(x_2))^n e^{-\rho M_{H}^{(E)}(x_{2})} \textrm{d} x_2 \textrm{d} y_2 \\
&= L \sqrt{\frac{\beta}{\pi}}  \frac{\Gamma\pr{m+1,  \frac{\rho\pi}{2\beta} } }{\rho^{m+1}}
\label{edge}.}

The corner contribution to~\eqref{inta} can be obtained by considering a wedge domain $\mathcal{V}=\{(r,\theta) : \theta\in(0,\phi) \}$ in polar coordinates for general angle $\phi<\pi$.
Expanding $H(\textbf{r}_{12})$ to linear order in $r_2\approx 0$, i.e. near the corner we get
\est{
M_{H}^{(C)} 
&= \! \int_{0}^{\phi} \! \int_{0}^{\infty} \!\! r_1 e^{-\beta r_{1}^{2}}(1+2 \beta r_1 r_2 \cos (\theta_1-\theta_2)) \textrm{d} r_1 \textrm{d} \theta_1 \\
&= \frac{\phi}{2 \beta} +  \sqrt{\frac{\pi}{\beta}} \frac{\sin(\phi-\theta_2) +\sin\theta_2}{2} r_2 +\mathcal{O}(r_2^2)
,}
\es{
I_{m}^{(C)}&= \sum_{n=0}^{m}  \binom {m} {n}  \pr{\frac{\phi}{2\beta}}^{m-n}\frac{4 \beta \Gamma(n+2)}{ \pi \sin\phi \rho^{n+1}} e^{-\rho \frac{\phi}{2\beta}}
\label{corner}.
}
Note that $I_{m}^{(C)}$ can be expressed as a single term but does not provide further insight and so is left as such.

%\subsubsection{3D Domains}

We now repeat the above calculations for  $\mathcal{V}\subset \R^3$. 
Therefore, $I_m$ now decomposes into a homogeneous bulk contribution ($B$), a surface area contribution ($S$), and edge ($E$) and corner ($C$) contributions
\es{
I_m \approx I_m^{(B)} + I_m^{(S)} + \sum_{E_i} I_m^{(E_i)} + \sum_{C_i} I_m^{(C_i)}
\label{inta2}
.}
We will restrict the discussion to  domains belonging to the set of right prisms; a polyhedron that accurately models many geometries that can be found in practice e.g. many room
configurations in a modern building.
As there is only one variable angle to consider we can benefit from the use of cylindrical coordinates when calculating $I_{m}^{(E)}$ and $I_{m}^{(C)}$.

Using spherical coordinates, we expand $H_{12}$ for $r_2\approx 0$ and find that for homogeneous domains $M_{H}^{(B)} \!\! = \frac{\pi^{3/2}}{\beta^{3/2}}$, and $I_{m}^{(B)}= V \pr{\frac{\pi}{\beta}}^{3m/2} e^{-\rho \pr{\frac{\pi}{\beta}}^{3/2}}$.

For the surface area contribution, we expand $H_{12}$ to linear order in $r_2\approx R$, where $R=\sqrt{S/(4\pi)}$ and $S$ is the total surface area of the domain. In doing so we are effectively saying that the surface area contribution to \eqref{int} for an arbitrary right prism is equal to that of a sphere of equal surface area. 
Details such as corners and edges are ignored at this stage as they will be considered separately at a later stage. We find that $M_{H}^{(S)} \!\!\approx \frac{\pi^{3/2}}{2\beta^{3/2}} +\frac{\pi}{\beta}(R-r_2)$, %leading to
\es{
I_{m}^{(S)}= S \frac{\beta}{\pi \rho^{1+m}} \Gamma\pr{m+1,\frac{\rho \pi^{3/2}}{2\beta^{3/2}}}
.}

Where two planes meet at an angle $\phi\in(0,\pi)$, an edge of length $L$ is formed. To calculate the contribution to \eqref{int} due to the edge we express $H_{12}$ in cylindrical coordinates such that the edge is centred along the $z$-axis. We expand $H_{12}$ about $r_2 \!\approx \!z_2 \!\approx\! 0$, keeping only linear terms we get $M_{H}^{(E)} = \frac{\sqrt{\pi} \phi}{2\beta^{3/2}} +
\frac{\pi}{2\beta}(\sin(\phi -\theta_2) + \sin\theta_2) r_2 + \mathcal{O}(r_2^{3})$,
\es{
%M_{H}^{(E)} 
%&= \int_{0}^{\phi}\int_{-\infty}^{\infty} \int_{0}^{\infty}  r_1 H_{12} \textrm{d} r_1 \textrm{d} z_1 \textrm{d} \theta_1 \\
%&= \frac{\sqrt{\pi} \phi}{2\beta^{3/2}} +
%\frac{\pi}{2\beta}(\sin(\phi -\theta_2) + \sin\theta_2) r_2 + %\mathcal{O}(r_2^{3}),\\
I_m^{(E)}&=L \! \sum_{n=0}^{m} \binom {m} {n} \!\pr{\frac{\sqrt{\pi} \phi}{2\beta^{3/2}}}^{m-n} \!\! \frac{4 \beta^2 \Gamma(2+n)}{\pi^2 \sin\phi \rho^{n+2}} e^{-\rho\frac{\sqrt{\pi}\phi}{2\beta^{3/2}}}
.}

Finally, corners in right prisms are formed where 3 edges come together, 2 of which are at an angle $\phi$ and the 3rd is perpendicular to both. We center the corner at the origin with the perpendicular edge running along the positive $z$-axis. We expand $H_{12}$ about $r_2 \approx z_2 \approx 0$ keeping only linear terms and calculate $M_{H}^{(C)} =
\frac{\sqrt{\pi} \phi}{4\beta^{3/2}} + \frac{ \phi}{2\beta}z_2 +
\frac{\pi}{4\beta}(\sin(\phi -\theta_2) + \sin\theta_2) r_2$
\es{
%M_{H}^{(C)} 
%= \int_{0}^{\phi}\int_{0}^{\infty} \int_{0}^{\infty}  r_1 H_{12} \textrm{d} r_1 \textrm{d} z_1 \textrm{d} \theta_1 \\
%&=
%\frac{\sqrt{\pi} \phi}{4\beta^{3/2}} + \frac{ \phi}{2\beta}z_2 +
%\frac{\pi}{4\beta}(\sin(\phi -\theta_2) + \sin\theta_2) r_2
%,\\
I_m^{(C)}&= \sum_{n=0}^{m} \binom {m} {n} \pr{\frac{\sqrt{\pi}\phi}{4 \beta^{3/2}}}^{m-n}
\frac{16\beta^{3}\Gamma(n+3)}{\pi^2 \phi \sin\phi \rho^{n+3}} e^{-\rho \frac{\sqrt{\pi} \phi}{4 \beta^{3/2}}}
.
}

In hindsight of the above calculations, it easy to see that 
2D and 3D contributions to \eqref{int} are similar in structure, hence hinting towards possible generalization to arbitrary dimensions $d>0$.
Furthermore, we observe that the dominant contribution at hight densities $\rho$ comes from $I_{m}^{(C)}$. Physically, this makes sense as $P_{md}(k)$ is most probable to fail near the hardest to connect to region of $\mathcal{V}$, i.e. the sharpest corner, in our case characterised by its angle $\phi$.

\section{Global Approach}
\label{sec:kconn}

\begin{figure}[t]
\begin{center}
\includegraphics[scale=0.176]{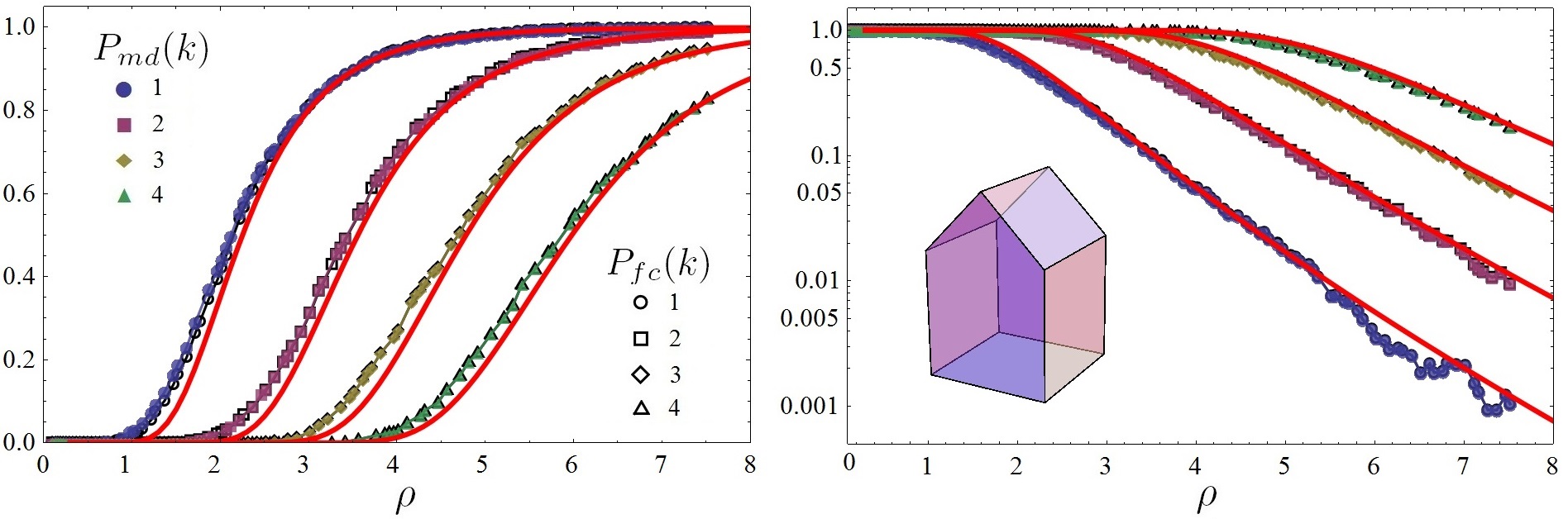}
\caption{\label{fig:soft}
\textit{Left:} Computer simulation of $P_{md}(k)$ (filled markers) and $P_{fc}(k)$ (hollow markers) for $k\in[1,4]$ using $\beta=1$ in a 3D house domain of sides $L=5$ and $L/\sqrt{2}$. The thick red curves are the analytic approximation of~\eqref{Pmd}. \textit{Right:} $1-P_{md}(k)$ and $1-P_{fc}(k)$ on a log-linear scale.
}
\end{center}
\end{figure}

For a given configuration of node positions we define $P_{fc}(\mathbf{r}_{1},\ldots, \mathbf{r}_{N},k)$ as the probability of the corresponding network to be $k$-connected. The average of this quantity over all possible node configurations $P_{fc}(k)=\langle  P_{fc}(\mathbf{r}_{1},\ldots, \mathbf{r}_{N},k) \rangle$ is the overall probability of obtaining a $k$-connected network.
It is clear that a $k$-connected network has minimum degree $k$. 
The opposite is not true however and hence the former set is a subset of the latter and so $P_{fc}(k)\leq P_{md}(k)$. 
For instance a network consisting of two pairs of connected nodes has minimum degree $1$ but is not $1$-connected.
Nevertheless, the two concepts are strongly correlated, particularly in the high density limit where the two converge ~\cite{penrose1999k,bettstetter2002minimum}.
Indeed, in Fig.~\ref{fig:soft}, $P_{fc}(k)$ (shown in hollow markers) follows $P_{md}(k)$ very closely from below.
Understanding the subtle differences between $P_{fc}(k)$ and $P_{md}(k)$ has posed a difficult challenge to the graph theoretic community since the early 80's and has ever since been approached from a variety of different directions. Here, through simple argumentation and the use of a cluster expansion for $P_{fc}(1)$ deriving from statistical physics \cite{coon2012full}, we will show that $P_{fc}(k)$ and $P_{md}(k)$ have the same asymptotic distribution.

%\subsection{A cluster expansion approach}

The probability that nodes connect (or not) leads to the trivial identity $1\equiv H_{ij}+ (1-H_{ij})$.
Multiplying over all possible links with nodes in $S$, expresses the probability of all possible combinations.
This can be written as
\es{
1= \prod_{(i,j)\in S^2; i<j}[H_{ij}+ (1-H_{ij})]= \sum_{g\in G^S}\mathcal{H}_{g}.
\label{2}}
where $\mathcal{H}_{g}=\prod_{(i,j)\in g}H_{ij}\prod_{(i,j)\not\in g}(1-H_{ij})$. Recall that $g=(S,L)$ is a network consisting of the set of nodes $S=\{1,2,3,\ldots,N\}$ paired by the collection $L\subseteq \{(i,j)\in S^2 :i<j\}$ of direct links.
As a slight abuse of notation we have used
$(i,j)\in g$ to denote that $(i,j)$ is an element of the set of links $L$ associated with $g$.

The sum in equation~\eqref{2} contains $2^{N(N-1)/2}$ separate terms and can be expressed as collections of terms determined by their largest cluster:
\es{
\label{3} 1=\sum_{g\in G^{S}_N}\mathcal{H}_{g}+ \sum_{g\in G^{S}_{N-1}}\mathcal{H}_{g} + \ldots +\sum_{g\in G^{S}_{1}}\mathcal{H}_{g},
}
where $G^S$ is the set of graphs with nodes in $S$, and $G^S_j$ the set of graphs with nodes in
$S$ and largest connected component (cluster) of size $j$ with $1\leq j\leq N$.
We identify the first term on the RHS of \eqref{3} as the probability of the associated network being $1$-connected $P_{fc}(\mathbf{r}_{1},\ldots, \mathbf{r}_{N},1)$.
Hence, rearranging equation~\eqref{3} and averaging over all possible node configurations we get
\es{
P_{fc}(1)= 1- \langle\!\!\!   \sum_{g\in G_{N-1}^{S} }
\!\!\! \mathcal{H}_{g}  \rangle
-\langle\!\!\!  \sum_{g\in G_{N-2}^{S} }
\!\!\! \mathcal{H}_{g}  \rangle -\ldots
\label{pfc1}
.}
Equation~\eqref{pfc1} clearly confirms the physical picture that at high densities, full connectivity is simply the complement of the probability of an isolated node i.e. a node of degree $0$.
Moreover, second order corrections are due to scenarios involving a single cluster of size $N-2$. 
At high densities it was shown that \eqref{pfc1} is given by 
$P_{fc}(1)= 1-\rho \int_{\mathcal{V}} e^{-\rho M_{H}\textbf{r}_1} 
\textrm{d}\textbf{r}_1$ \cite{coon2012full}.

%\subsection{\textit{k}-connectivity at high node densities}

The probability of a network to be $2$-connected can be expressed as $P_{fc}(2)= P_{fc}(1)- X(1)$ where $X(1)\! \geq \!0$ is the probability of obtaining a fully connected network which is not $2$-connected. This follows from the fact that $k$-connectivity implies $(k-1)$-connectivity. 
At high densities, a fully connected network which is not $2$-connected will typically contain a single node which is of degree one
\es{
X(1)&\approx \langle \sum_{i=1}^{N} \sum_{j\not= i} H_{ij} \prod_{k\not=j\not=i}(1-H_{ik }) \rangle \\
%&= N(N-1) \langle H_{12} \prod_{k=3}^{N}(1-H_{1,k}) \rangle \\
%&= \frac{N(N-1)}{V^{2}}\int_{\mathcal{V}} M_{H}(\mathbf{r}_1) \pr{1-\frac{1}{V} H_{M}(\textbf{r}_{1})}^{N-2} \textrm{d} \textbf{r}_{1}\\
&= \rho^{2}\int_{\mathcal{V}} M_{H}(\mathbf{r}_1) e^{-\rho H_{M}(\textbf{r}_{1})} \textrm{d} \textbf{r}_{1}
,}
for $N\gg 1$. Repeating the same argument $k$ times we get
\es{
P_{fc}(k)&= P_{fc}(1)- \sum_{m=1}^{k-1}X(m) = 1- \sum_{m=0}^{k-1}X(m),
\\
X(m)&= \frac{\rho^{m+1}}{m!}\int_{\mathcal{V}} M_{H}^{m} (\textbf{r}_1) e^{-\rho M_{H} (\textbf{r}_1)}
\textrm{d}  \textbf{r}_1
,
\label{fc}
}
where $X(m)$ and is the probability of obtaining an $m$-connected network which is not $(m+1)$-connected. 
Examples of networks where a single node prohibits $(k+1)$-connectivity can be seen in Fig.~\ref{fig:123conn} for $k=1,2,3$.
Finally, noting that $(1-x)^N \sim 1-Nx$ for $N\gg1$ and $x<1$, and comparing equation \eqref{Pmd} to \eqref{fc}, we can conclude that $P_{fc}(k)$ and $P_{md}(k)$ have the same asymptotic distribution in $\rho$.
Moreover, $k$-link-connectivity, where a network remains connected whenever fewer than $k$ links are removed, is sandwiched by $P_{md}(k)$ and $P_{fc}(k)$.

\section{Numerical verification}

We numerically test our results in a 3D house domain as seen in Fig.~\ref{fig:soft} with sides $L=5$ and $L/\sqrt{2}$ (using $\beta=1$).
The left panel of Fig.~\ref{fig:soft} shows in full markers the numerically obtained $P_{md}(k)$ for $k\in[1,4]$, while the S-shaped red curves is the analytic approximation of~\eqref{Pmd}.
An excellent agreement is observed, especially for high densities as can be seen in the right panel of Fig.~\ref{fig:soft} depicting $1-P_{md}(k)$ on a log-linear scale. 
The hollow markers correspond to $P_{fc}(k)$ and closely follow $P_{md}(k)$ from below.
At high densities, the difference becomes increasingly difficult to make, confirming that the two observables have the same asymptotic distribution as $\rho\to\infty$.
We attribute the difference between theory and simulations at low densities to the Poisson approximation of \eqref{nd} and the independence assumption in \eqref{pmd}.
Finally, while the theoretical curve (in red) is systematically lower than both simulation results at low densities, it overtakes $P_{fc}(k)$ and better approximates $P_{md}(k)$ at medium-to-high densities (subject to small fluctuations) as expected.

%%%%%%%%%%%%%%%%%%%%%%%%%%

\section{Conclusions}
\label{sec:concl}

We have investigated the probability of forming a $k$-connected random network $P_{fc}(k)$ confined within convex 2 and 3 dimensional domains and have found that for probabilistic link models, $P_{fc}(k)$ is governed by
boundary effects due to distinct microscopic details of the network domain such as sharp corners and edges which can be singled-out and analysed independently.
As a result, we have obtained accurate approximations for contributions to $P_{fc}$ due to the bulk, surface area, edges and corners of the domain. 
These contributions can now be easily calculated for an arbitrarily complex (but convex) domain, and summed up to give accurate predictions to $P_{fc}(k)$.
We have confirmed the validity of our results through computer simulations in a three dimensional house domain.

The results presented here can have direct and applicable benefits in the design of wireless multi-hop relay networks where communication devices (nodes) pass messages to each other without the need of a central router.
Significantly, our analysis enables network engineers and researchers to glean important information that will dictate how optimal deployments can be made in practice e.g., for wireless ad hoc vehicular and sensor networks \cite{Brendin10}. 
One timely example that is currently receiving a considerable amount of attention in Europe is the `smart meter roll-out' (see European commission mandate M/441~\cite{M441}). 
Such networks, aimed at supporting the so-called smart grid, are typically random and dense at a local (e.g., neighbourhood) level, and require a high degree of resilience to node failures owing to the significance of their role in smart grid operation. 
Consequently, one may consider ways to mitigate microscopic boundary effects by e.g. increasing the signal power or number of communication channels.

Finally, our work is not however restricted to communication networks and can provide further insight on the difficult problem of resilience reliability and control \cite{Nepusz12} of large and highly interconnected real networks. Example applications of our theoretical work may include systems of water, food and fuel supply, financial transactions \cite{palla2007quantifying,Gai10,Vitali11} and power transmission \cite{Mott13}, or smaller, boundary dominated ones  involving for instance the electrical conductivity of carbon nano-tubes \cite{kyrylyuk2011controlling}.

%%%%%%%%%%%%%%%%%%%%%%%%%%

\acknowledgments
The authors thank the directors of the Toshiba Telecommunications Research Laboratory for their support.

%%%%%%%%%%%%%%%%%%%%%%%

\end{document}